\documentstyle{article}

\newcommand{\ba}{\begin{eqnarray}}
\newcommand{\ea}{\end{eqnarray}}
\input{psfig}
\def\ii{\'{\i}}
\begin{document}
\pagestyle{plain}

\title{Algebraic model of an oblate top}
\author{R. Bijker\\
Instituto de Ciencias Nucleares, U.N.A.M., A.P. 70-543,\\
04510 M\'exico, D.F., M\'exico
\and
A. Leviatan\\
Racah Institute of Physics, The Hebrew University,\\
Jerusalem 91904, Israel} 
\maketitle

\begin{abstract}
We consider an algebraic treatment of a three-body system.  
In particular, we develop the formalism 
for a system of three identical objects and discuss an 
application to nonstrange baryon resonances which are interpreted 
as vibrational and rotational excitations of an oblate symmetric top. 
We derive closed expressions for a set of elementary form factors 
that appear in the calculation of both electromagnetic, strong and 
weak couplings of baryons.
\end{abstract}
\vspace{2cm}
\begin{center}
Invited talk at `Symmetries in Science IX',\\
Bregenz, Austria, August 6-10, 1996\\
Preprint: nucl-th/9609014 
\end{center}

\clearpage
\section{Introduction}
\setcounter{equation}{0}

The development of spectrum generating algebras and the study of 
exactly solvable systems have played an important role in all 
fields of physics \cite{SGA}. In particular, in spectroscopic studies 
algebraic methods are very useful to study the symmetries and selection 
rules, to classify the basis states, and to calculate matrix elements.
An exactly solvable model that is of special interest is the symmetric 
top. As an example, we mention a study by Bohm and Teese \cite{BT}  
in which the rotational spectrum of a prolate symmetric top was 
treated in terms of $SO(3,2)$ with representation doubling by parity. 

The aim of this contribution is to study a realization of a symmetric 
top in which all vibrational and rotational degrees of freedom are 
present from the outset. A good candidate for such an approach is 
provided by the so-called algebraic method 
that was developed for the collective vibrations and rotations  
of nuclei \cite{ibm}, and later extended to rovibrational 
states of molecules \cite{vibron,thebook,BDL} and collective excitations 
of the nucleon \cite{BIL}. This approach is based on the 
general criterion \cite{FI} to take $U(k+1)$ as a spectrum generating 
algebra for a bound-state problem with $k$ degrees of freedom and 
assigning all states to the symmetric representation $[N]$ of $U(k+1)$. 
For collective nuclei this led to the introduction of the $U(6)$ 
interacting boson model \cite{ibm} and for diatomic molecules to the 
$U(4)$ vibron model \cite{vibron}.

In this contribution we apply the algebraic method to a three-body 
system. The dynamics of this system is determined by the six degrees 
of freedom of the two relative coordinates, which in the algebraic 
approach leads to a $U(7)$ spectrum generating algebra. 
We develop the formalism for a system of three identical 
objects and discuss an application to nonstrange baryon resonances which 
are interpreted as vibrations and rotations of an oblate symmetric top. 
 
\section{Algebraic treatment of a three-body system}
\setcounter{equation}{0}

The geometry of a three-body system is characterized by two 
relative coordinates which we choose as the relative Jacobi coordinates 
\ba
\vec{\rho} &=& \frac{1}{\sqrt{2}} (\vec{r}_1 - \vec{r}_2) ~,
\nonumber\\
\vec{\lambda} &=& \frac{1}{\sqrt{6}} (\vec{r}_1 + \vec{r}_2 -2\vec{r}_3) ~. 
\label{jacobi}
\ea
Here $\vec{r}_1$, $\vec{r}_2$ and $\vec{r}_3$ are the coordinates of 
the three objects. Instead of using coordinates and momenta we prefer 
to use a second quantized formalism, in which we introduce a dipole 
boson for each independent relative coordinate and an auxiliary scalar 
boson
\ba 
b^{\dagger}_{\rho,m} ~, \; b^{\dagger}_{\lambda,m} ~, \; 
s^{\dagger} \hspace{1cm} (m=-1,0,1) ~. \label{bb}
\ea
The scalar boson does not represent an independent degree of freedom, 
but is added under the restriction that the total number of bosons 
$N=n_{\rho}+n_{\lambda}+n_s$ is conserved. This procedure leads to a 
compact spectrum generating algebra of $U(7)$ whose 49 generators are 
\ba
b^{\dagger}_{i,m} b_{j,m^{\prime}} ~, \;\;\; 
b^{\dagger}_{i,m} s ~, \;\;\; 
s^{\dagger} b_{i,m} ~, \;\;\; s^{\dagger} s ~,
\ea
with ($m,m^{\prime}=-1,0,1$) and ($i,j=\rho,\lambda$).
The introduction of the scalar boson is 
just an elegant and efficient way by means of which the full dynamics 
of two vectors can be investigated. This includes situations in which 
there is a strong mixing of the oscillator basis (collective models).
For a system of interacting bosons the model space is spanned by the 
symmetric irreps $[N]$ of $U(7)$, which contains the oscillator shells 
with $n=n_{\rho}+n_{\lambda}=0,1,2,\ldots, N$. The value of $N$ determines
the size of the model space. 

For three identical objects ({\it e.g.} for X$_3$ molecules or 
nonstrange qqq baryons) the Hamiltonian (or mass operator) has to be 
invariant under the permuation group $S_3$. 
The permutation symmetry of three identical objects is determined 
by the transposition $P(12)$ and the cyclic permutation $P(123)$ \cite{KM}.
All other permutations can be expressed in terms of these two 
elementary ones. Algebraically, these operators can be expressed in terms 
of the generators $G_{ij}=\sum_{m} b^{\dagger}_{i,m} b_{j,m}$ 
that act in index space ($i,j=\rho,\lambda$). We find 
\ba
P(12) &=& 
\mbox{exp} \left\{ -i \, \frac{\pi}{2} (\hat F_3 + \hat n) \right\} 
\;=\; \mbox{exp} \left\{ -i \, \pi \hat n_{\rho} \right\} ~,
\nonumber\\
P(123) &=& \mbox{exp} \left\{ -i \, \frac{2\pi}{3} \hat F_2 \right\} ~. 
\label{point}
\ea
The operators $\hat F_2$ and $\hat F_3$ are two components of the 
$SU(2)$ pseudo-spin in index space, which in turn is 
related to the vortex spin \cite{Rowe} 
\ba
\hat F_1 &=& \sum_{m} \left( b^{\dagger}_{\rho,m} b_{\lambda,m} 
+ b^{\dagger}_{\lambda,m} b_{\rho,m}  \right) ~,
\nonumber\\
\hat F_2 &=& -i \sum_{m} \left( b^{\dagger}_{\rho,m} b_{\lambda,m} 
- b^{\dagger}_{\lambda,m} b_{\rho,m}  \right) ~,
\nonumber\\
\hat F_3 &=& \sum_{m} \left( b^{\dagger}_{\rho,m} b_{\rho,m} 
- b^{\dagger}_{\lambda,m} b_{\lambda,m} \right) ~. \label{fspin}
\ea
The operator $\hat n$ counts the total number of dipole bosons 
\ba
\hat n &=& \hat n_{\rho} + \hat n_{\lambda} \;=\; \sum_{m} \left(
  b^{\dagger}_{\rho,m} b_{\rho,m} 
+ b^{\dagger}_{\lambda,m} b_{\lambda,m} \right) ~. \label{nph}
\ea
Since the permutation group $S_3$ is isomorphic to $D_3$, one can 
either use the irreducible representations of 
$S_3$ or those of $D_3$ to label the three symmetry classes. 
In this contribution we use the $D_3$ labeling: 
$A_1$ and $A_2$ for the one-dimensional symmetric and antisymmetric 
representations, and $E$ for the two-dimensional representation.

All operators of interest are expressed in terms of the building 
blocks of Eq.~(\ref{bb}) which transform under $D_3$ as $E_{\rho}$, 
$E_{\lambda}$ and $A_1$, respectively. 
With the help of Eq.~(\ref{point}) one can 
construct physical operators with the appropriate symmetry properties. 
The most general form of the Hamiltonian, that preserves angular momentum 
and parity, transforms as a scalar ({\it i.e.} $A_1$) under $D_3$ and 
contains at most two-body interactions is given by 
\ba
\hat H &=& H_0 + \epsilon_s \, s^{\dagger} \tilde{s} 
- \epsilon_p \, (b_{\rho}^{\dagger} \cdot \tilde{b}_{\rho}
+ b_{\lambda}^{\dagger} \cdot \tilde{b}_{\lambda})
+ u_0 \, (s^{\dagger} s^{\dagger} \tilde{s} \tilde{s}) - u_1 \,
  s^{\dagger} ( b^{\dagger}_{\rho} \cdot \tilde{b}_{\rho}
+ b^{\dagger}_{\lambda} \cdot \tilde{b}_{\lambda} ) \tilde{s}
\nonumber\\
&& + v_0 \, \left[ ( b^{\dagger}_{\rho} \cdot b^{\dagger}_{\rho}
+ b^{\dagger}_{\lambda} \cdot b^{\dagger}_{\lambda} ) \tilde{s} \tilde{s} 
+ s^{\dagger} s^{\dagger} ( \tilde{b}_{\rho} \cdot \tilde{b}_{\rho}
+ \tilde{b}_{\lambda} \cdot \tilde{b}_{\lambda} ) \right]
+ c_1 \, ( b^{\dagger}_{\rho} \times b^{\dagger}_{\lambda} )^{(1)} 
\cdot ( \tilde b_{\lambda} \times \tilde b_{\rho} )^{(1)}
\nonumber\\
&& + \sum_{l=0,2} c_l \, \left[ 
( b^{\dagger}_{\rho} \times   b^{\dagger}_{\rho} 
- b^{\dagger}_{\lambda} \times b^{\dagger}_{\lambda} )^{(l)} \cdot
( \tilde{b}_{\rho} \times \tilde{b}_{\rho}
- \tilde{b}_{\lambda} \times \tilde{b}_{\lambda} )^{(l)}
+ 4 \, ( b^{\dagger}_{\rho} \times b^{\dagger}_{\lambda})^{(l)} \cdot
       ( \tilde b_{\lambda} \times \tilde b_{\rho})^{(l)} \right]
\nonumber\\
&& + \sum_{l=0,2} w_l \, ( b^{\dagger}_{\rho} \times b^{\dagger}_{\rho}
  + b^{\dagger}_{\lambda} \times b^{\dagger}_{\lambda} )^{(l)} \cdot
  ( \tilde{b}_{\rho} \times \tilde{b}_{\rho}
  + \tilde{b}_{\lambda} \times \tilde{b}_{\lambda} )^{(l)} ~.
\label{ms3}
\ea
Here $\tilde{b}_{\rho,m} = (-1)^{1-m} b_{\rho,-m}$~,
$\tilde{b}_{\lambda,m} = (-1)^{1-m} b_{\lambda,-m}$ 
and $\tilde{s} = s$~. The dots indicate scalar products and the 
crosses tensor products. 
The Hamiltonian of Eq.~(\ref{ms3}) contains several interesting 
limiting situations. The harmonic oscillator model arises for 
$v_0=0$, {\it i.e.} no coupling between different harmonic oscillator 
shells. This scenario corresponds to the $U(7) \supset U(6)$ 
group reduction. On the other hand, for the choice $v_0 \neq 0$ 
the eigenfunctions are collective in nature, since they are spread 
over many oscillator shells. 

The eigenvalues and corresponding eigenvectors can be obtained exactly 
by diagonalization in an appropriate basis. 
The wave functions obtained in this way have by construction good 
angular momentum $L$, parity $P$, and permutation symmetry $t$. 

\section{Permutation symmetry}
\setcounter{equation}{0}

The permutation symmetry of a given wave function can be determined 
from the transformation properties under $P(12)$ and $P(123)$ \cite{KM}.
Since the Hamiltonian of Eq.~(\ref{ms3}) is invariant 
under the transposition $P(12)$, basis states with $n_{\rho}$ even and 
$n_{\rho}$ odd do not mix, and can therefore be treated separately. 
This allows one to distinguish wave functions with $t=A_1$ or $E_{\lambda}$ 
from wave functions with $t=A_2$ or $E_{\rho}$
\ba
P(12) \left( \begin{array} {l} | \psi_{A_1} \rangle \\ 
| \psi_{A_2} \rangle \\ | \psi_{E_{\rho}} \rangle \\ 
| \psi_{E_{\lambda}} \rangle \end{array} \right) &=&
\left( \begin{array}{rrrr} 1 & 0 & 0 & 0 \\ 0 & -1 & 0 & 0 \\
0 & 0 & -1 & 0 \\ 0 & 0 & 0 & 1 \end{array} \right)
\left( \begin{array} {l} | \psi_{A_1} \rangle \\ 
| \psi_{A_2} \rangle \\ | \psi_{E_{\rho}} \rangle \\ 
| \psi_{E_{\lambda}} \rangle \end{array} \right) ~. \label{P12}
\ea
The cyclic permutation $P(123)$ can be used to distinguish 
wave functions with $t=A_1$ or $A_2$ from wave functions 
with $t=E_{\lambda}$ or $E_{\rho}$ 
\ba
P(123) \left( \begin{array} {l} | \psi_{A_1} \rangle \\ 
| \psi_{A_2} \rangle \\ | \psi_{E_{\rho}} \rangle \\ 
| \psi_{E_{\lambda}} \rangle \end{array} \right) &=&
\left( \begin{array}{cccc} 1 & 0 & 0 & 0 \\ 0 & 1 & 0 & 0 \\
0 & 0 &  \cos (2\pi/3) & -\sin (2\pi/3) \\ 
0 & 0 &  \sin (2\pi/3) &  \cos (2\pi/3) \end{array} \right)
\left( \begin{array} {l} | \psi_{A_1} \rangle \\ 
| \psi_{A_2} \rangle \\ | \psi_{E_{\rho}} \rangle \\ 
| \psi_{E_{\lambda}} \rangle \end{array} \right) ~. \label{P123}
\ea 
In Eqs.~(\ref{P12}) and~(\ref{P123}) we have used that the  
application of $P(12)$ and $P(123)$ on $| \psi_t \rangle$ 
means carrying out the inverse operation on the boson operators, 
$b^{\dagger}_{\rho,m}$ and $b^{\dagger}_{\lambda,m}$ \cite{KM,Wigner}. 
In practice, the wave functions $| \psi_{E_{\rho}} \rangle$ and 
$| \psi_{E_{\lambda}} \rangle$ are obtained from separate 
diagonalizations, and hence are determined up to a sign. 
Eq.~(\ref{P123}) can be used to 
determine their relative sign, so that they transform as the two 
components of the mixed symmetry doublet with $t=E$ symmetry. 

In addition to angular momentum, parity and permutation symmetry, 
$L^P_t$, the Hamiltonian of Eq.~(\ref{ms3}) has another symmetry. 
Since $\hat H$ commutes with the operator $\hat F_2$ of 
Eq.~(\ref{fspin}), its eigenstates can also be labeled by the 
eigenvalues of $\hat F_2$: $m_F=0,\pm 1,\ldots, \pm N$. We denote 
these eigenstates by $| \phi_{m_F} \rangle$~. 
Since the transposition $P(12)$ anticommutes with $\hat F_2$
\ba
P(12) \, \hat F_2 &=& - \hat F_2 \, P(12) ~, \label{p12f2}
\ea
the simultaneous eigenfunctions of $P(12)$ and $\hat H$ are given 
by the linear combinations
\ba
| \psi_1 \rangle &=& \frac{-i}{\sqrt{2(1+\delta_{m_F,0})}} \, 
\left[ | \phi_{+m_F} \rangle - | \phi_{-m_F} \rangle \right] ~,
\nonumber\\
| \psi_2 \rangle &=& \frac{(-1)^{\nu}}{\sqrt{2(1+\delta_{m_F,0})}} \, 
\left[ | \phi_{+m_F} \rangle + | \phi_{-m_F} \rangle \right] ~. 
\label{wfp12}
\ea
Here we have introduced the label $\nu$ by $m_F=\nu$ (mod 3). 
These wave functions satisfy 
\ba
P(12) \left( \begin{array}{c} | \psi_1 \rangle \\ 
| \psi_2 \rangle \end{array} \right) 
&=& \left( \begin{array}{rr} -1 & 0 \\ 0 & 1 \end{array} \right)
\left( \begin{array}{c} | \psi_1 \rangle \\ | \psi_2 \rangle 
\end{array} \right) ~.
\ea
The cyclic permutation $P(123)$ acts in Fock space as a rotation 
generated by $\hat F_2$ (see Eq.~(\ref{point}))
\ba
P(123) \left( \begin{array}{c} | \psi_1 \rangle \\ 
| \psi_2 \rangle \end{array} \right) 
&=& \left( \begin{array}{cc} \cos(2 \pi m_F/3) 
& (-1)^{\nu} \sin(2 \pi m_F/3) \\
-(-1)^{\nu} \sin(2 \pi m_F/3) & \cos(2 \pi m_F/3) \end{array} \right)
\left( \begin{array}{c} | \psi_1 \rangle \\ 
| \psi_2 \rangle \end{array} \right) ~.
\ea
According to Eq.~(\ref{P123}) the wave functions $| \psi_1 \rangle$ 
($| \psi_2 \rangle$) transform for $\nu=0$ as $t=A_2$ ($A_1$), and 
for $\nu=1,2$ as $t=E_{\rho}$ ($E_{\lambda}$).
The special connection between the label $m_F=\nu$ (mod 3) and the 
permutation symmetry only holds for Hamiltonians that commute with 
$\hat F_2$. All one- and two-body $D_3$ invariant interactions indeed 
satisfy this property, and hence their eigenstates can be labeled by 
$M_F=|m_F|$ and $L^P_t$~. However, at the three-body level there are $D_3$ 
invariant interactions that mix 
states with $\Delta m_F=\pm 6$ \cite{BDL,Watson}. 
The permutation symmetry can then still be determined by the general 
procedure outlined in Eqs.~(\ref{P12}) and (\ref{P123}).

\section{Oblate symmetric top}
\setcounter{equation}{0}

An analysis of the equilibrium shape of the potential energy surface 
that corresponds to Eq.~(\ref{ms3}) yields that the only stable 
nonlinear configuration is that of an equilateral triangle \cite{BL7}. 
This equilibrium configuration is represented in $U(7)$ by an 
intrinsic (or coherent) state in the from of a condensate 
\ba
\frac{1}{\sqrt{N!}} ( b_c^{\dagger} )^N \, | 0 \rangle ~, \label{cond}
\ea
with
\ba
b_c^{\dagger} &=& \left[ s^{\dagger} + R ( b_{\rho,y }^{\dagger} 
+ b_{\lambda,x}^{\dagger} )/\sqrt{2} \right]/\sqrt{1+R^2} ~. \label{bc}
\ea
The equilibrium shape of an equilateral triangle is a result of the 
underlying $D_3$ symmetry. 

In order to analyze the vibrational and rotational excitations it is 
convenient to split the Hamiltonian into an intrinsic (vibrational) 
and a collective (rotational) part \cite{KL}. The intrinsic 
part of Eq.~(\ref{ms3}) by definition annihilates the condensate 
of Eqs.~(\ref{cond}) and~(\ref{bc}) and is given by \cite{BIL}
\ba
\hat H_{\mbox{int}} &=& \xi_1 \, \Bigl ( R^2 \, s^{\dagger} s^{\dagger}
- b^{\dagger}_{\rho} \cdot b^{\dagger}_{\rho}
- b^{\dagger}_{\lambda} \cdot b^{\dagger}_{\lambda} \Bigr ) \,
\Bigl ( R^2 \, \tilde{s} \tilde{s} - \tilde{b}_{\rho} \cdot \tilde{b}_{\rho}
- \tilde{b}_{\lambda} \cdot \tilde{b}_{\lambda} \Bigr )
\nonumber\\
&& + \xi_2 \, \Bigl [
\Bigl( b^{\dagger}_{\rho} \cdot b^{\dagger}_{\rho}
- b^{\dagger}_{\lambda} \cdot b^{\dagger}_{\lambda} \Bigr ) \,
\Bigl ( \tilde{b}_{\rho} \cdot \tilde{b}_{\rho}
- \tilde{b}_{\lambda} \cdot \tilde{b}_{\lambda} \Bigr )
+ 4 \, \Bigl ( b^{\dagger}_{\rho} \cdot b^{\dagger}_{\lambda} \Bigr ) \,
\Bigl ( \tilde{b}_{\lambda} \cdot \tilde{b}_{\rho} \Bigr ) \Bigr ] ~.
\label{oblate}
\ea
For the special case of $R^2=0$, the intrinsic Hamiltonian 
has $U(7) \supset U(6)$ symmetry and corresponds
to an anharmonic vibrator, whereas for $R^2=1$ and $\xi_2=0$ it has 
$U(7) \supset SO(7)$ symmetry and corresponds to a deformed 
oscillator. 

In the more general case with $R^2 \neq 0$ and $\xi_1$, $\xi_2>0$, 
the intrinsic Hamiltonian of Eq.~(\ref{oblate}) describes the 
vibrational excitations of an oblate symmetric top. 
This can be seen from a normal mode analysis. To leading order in $N$, 
$\hat H_{\mbox{int}}$ reduces to a harmonic form \cite{BDL,BIL}
\ba
\hat H_{\mbox{int}} &=& \epsilon_1 \, b^{\dagger}_{u} b_{u}
+ \epsilon_2 \, ( b^{\dagger}_{v} b_{v} + b^{\dagger}_{w} b_{w} ) ~,
\label{hvib} 
\ea
with eigenfrequencies $\epsilon_1 = 4 N \xi_1 R^2$ and 
$\epsilon_2 = 4 N \xi_2 R^2/(1+R^2)$~. The deformed bosons are 
given by 
\ba
b^{\dagger}_{u} &=& \left[ - R \, s^{\dagger} 
+ ( b_{\rho,y }^{\dagger} 
+ b_{\lambda,x}^{\dagger} )/\sqrt{2} \right]/\sqrt{1+R^2} ~, 
\nonumber\\
b^{\dagger}_{v} &=& ( b_{\lambda,x}^{\dagger} 
- b_{\rho,y}^{\dagger} )/\sqrt{2} ~,
\nonumber\\
b^{\dagger}_{w} &=& ( b_{\lambda,y}^{\dagger} 
+ b_{\rho,x}^{\dagger} )/\sqrt{2} ~. \label{bint}
\ea
The first term in Eq.~(\ref{hvib}) represents the symmetric stretching 
mode ($b_u$) and the second term a degenerate doublet of an antisymmetric 
stretching mode ($b_v$) and a bending mode ($b_w$). This is in agreement 
with the point-group classification of the fundamental vibrations 
of a symmetric $X_3$ configuration \cite{Herzberg}. Therefore, 
the deformed bosons of Eqs.~(\ref{bc}) and~(\ref{bint}) correspond to a 
geometry of an oblate symmetric top with the threefold symmetry axis 
along the $z$-axis.
In the large $N$ limit the vibrational spectrum is harmonic 
\ba
E_{\mbox{vib}} &=& \epsilon_1 \, v_1 + \epsilon_2 \, v_2 ~.
\ea

In a geometric description, the excitations of an oblate top 
are labeled by $(v_1,v_2^l);K,L^P_t,M$. Here $v_1$ denotes the number 
of quanta in the symmetric stretching mode which has $A_1$ symmetry, 
and $v_2$ the total number of quanta in the asymmetric 
stretching and the bending modes, 
which form a degenerate doublet with $E$ symmetry. The label $l$ is 
associated with the degenerate vibration. It is 
proportional to the vibrational angular momentum about the axis of 
symmetry and can have the values  
$l=v_2,v_2-2, \dots,1$ or 0 for $v_2$ odd or even, respectively.
The rotational states, which are characterized by the angular momentum 
$L$ and its projection $K$ on the three-fold symmetry axis,  
are arranged in bands built on top of each vibration. The projection 
$K$ can take the values $K=0,1,2,\dots$~, while the values of the 
angular momentum are $L=K,K+1,K+2,\ldots$~. 
The parity is $P=(-)^K$, $t$ denotes the transformation 
character of the total wave function under $D_3$, and $M$ is the 
angular momentum projection. For a given value of $l$ and $K$ 
the degeneracy of a state with angular 
momentum $L$ is given by $4(2L+1)/(1+\delta_{l,0})(1+\delta_{K,0})$. 

\section{Geometric interpretation of $M_F$}
\setcounter{equation}{0}

In section~3 we showed that the eigenstates of the algebraic 
Hamiltonian of Eq.~(\ref{ms3}) can be labeled by $M_F,L^P_t$~. 
The same holds for its intrinsic part, Eq.~(\ref{oblate}), which 
describes the vibrational excitations of an oblate top. 
In this section we wish to elucidate the role of the label $M_F$ 
in the context of the oblate symmetric top, and, in particular, 
its relation to the geometric labels $K$ and $l$.

\begin{figure}
\centering
\setlength{\unitlength}{1.0pt}
\begin{picture}(250,325)(0,20)
\thinlines
\put (  0, 20) {\line(1,0){250}}
\put (  0,325) {\line(1,0){250}}
\put (  0, 20) {\line(0,1){305}}
\put (250, 20) {\line(0,1){305}}
\thicklines
\put ( 30, 60) {\line(1,0){20}}
\put ( 30, 90) {\line(1,0){20}}
\put ( 30,150) {\line(1,0){20}}
\put ( 30,240) {\line(1,0){20}}
\put ( 80, 88) {\line(1,0){20}}
\put ( 80,148) {\line(1,0){20}}
\put ( 80,238) {\line(1,0){20}}
\put (130,142) {\line(1,0){20}}
\put (130,232) {\line(1,0){20}}
\put (180,222) {\line(1,0){20}}
\thinlines
\put ( 40,265) {$\vdots$} 
\put ( 30,290) {$M_F=0$}
\put ( 80,290) {$M_F=1$}
\put (130,290) {$M_F=2$}
\put (180,290) {$M_F=3$}
\put ( 35, 35) {$K$=0}
\put ( 85, 35) {$K$=1}
\put (135, 35) {$K$=2}
\put (185, 35) {$K$=3}
\put ( 52, 60) {$0^+_{A_1}$}
\put ( 52, 90) {$1^+_{A_2}$}
\put ( 52,150) {$2^+_{A_1}$}
\put ( 52,240) {$3^+_{A_2}$}
\put (102, 88) {$1^-_{E}$}
\put (102,148) {$2^-_{E}$}
\put (102,238) {$3^-_{E}$}
\put (152,142) {$2^+_{E}$}
\put (152,232) {$3^+_{E}$}
\put (202,222) {$3^-_{A_1A_2}$}
\end{picture}
\caption[Rotational spectrum of $(v_1,v_2^{l=0})$ band]
{\small Schematic representation of rotational spectrum of a 
$(v_1,v_2^{l=0})$ vibrational band. The states are labeled by $L^P_t$. 
Here $K$ denotes the absolute value of the projection of the angular 
momentum $L$ on the symmetry axis.
\normalsize}
\label{vib0}
\vspace{15pt}
\end{figure}
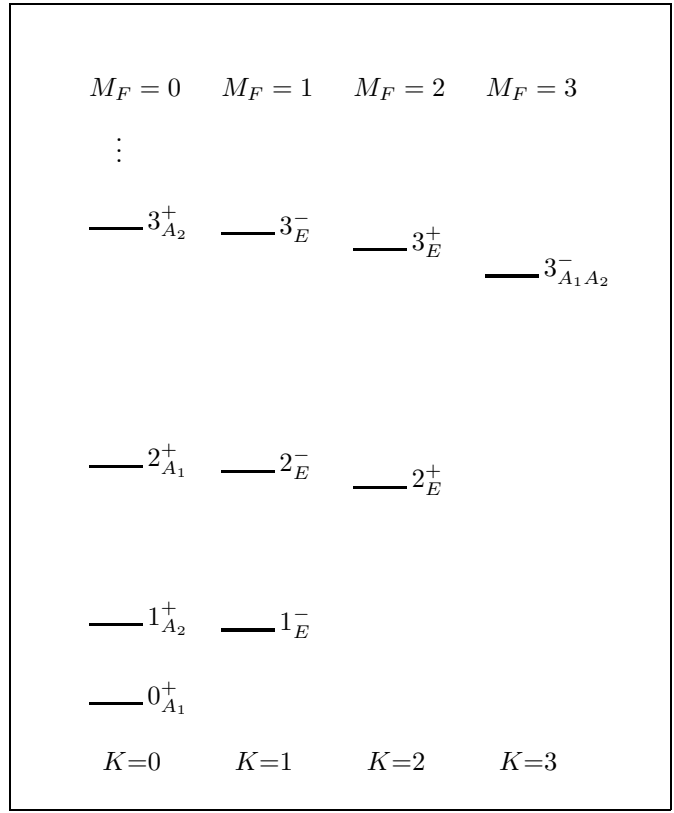

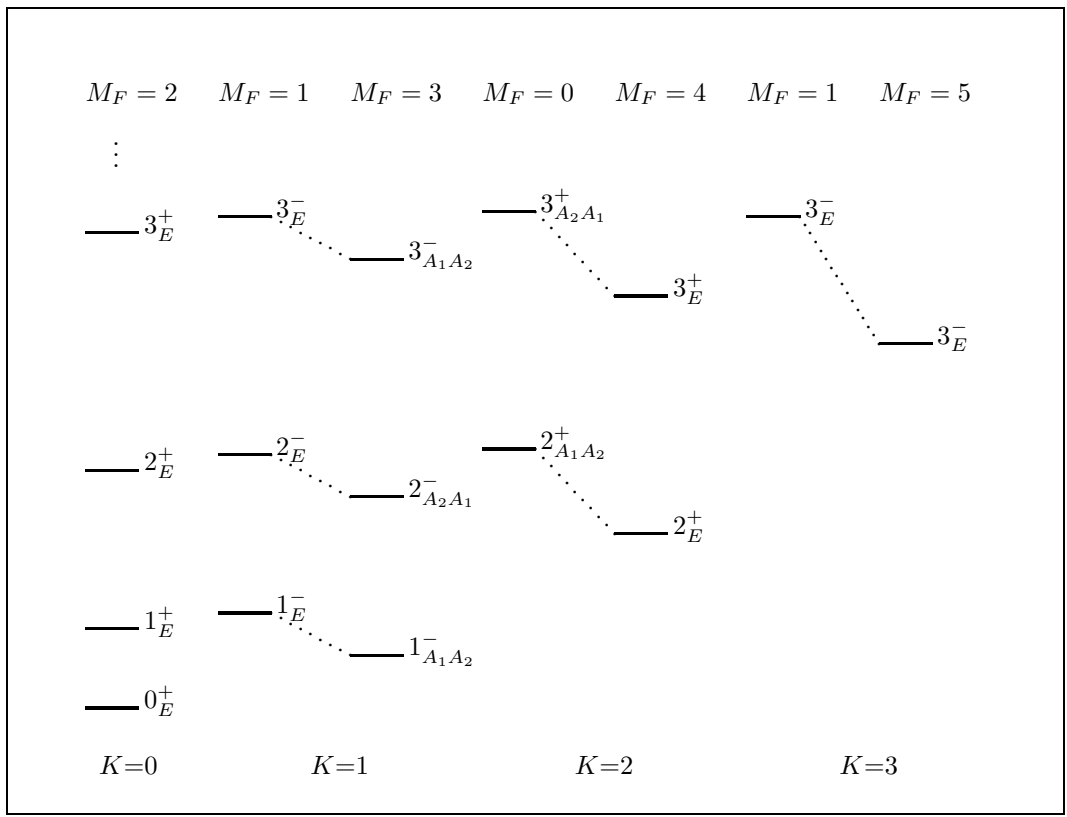
\begin{figure}
\centering
\setlength{\unitlength}{1.0pt}
\begin{picture}(400,325)(0,20)
\thinlines
\put (  0, 20) {\line(1,0){400}}
\put (  0,325) {\line(1,0){400}}
\put (  0, 20) {\line(0,1){305}}
\put (400, 20) {\line(0,1){305}}
\thicklines
\put ( 30, 60) {\line(1,0){20}}
\put ( 30, 90) {\line(1,0){20}}
\put ( 30,150) {\line(1,0){20}}
\put ( 30,240) {\line(1,0){20}}
\put ( 80, 96) {\line(1,0){20}}
\put ( 80,156) {\line(1,0){20}}
\put ( 80,246) {\line(1,0){20}}
\multiput (100, 96)(3.75,-2){9}{\circle*{0.1}}
\multiput (100,156)(3.75,-2){9}{\circle*{0.1}}
\multiput (100,246)(3.75,-2){9}{\circle*{0.1}}
\put (130, 80) {\line(1,0){20}}
\put (130,140) {\line(1,0){20}}
\put (130,230) {\line(1,0){20}}
\put (180,158) {\line(1,0){20}}
\put (180,248) {\line(1,0){20}}
\multiput (200,158)(3,-3.2){11}{\circle*{0.1}}
\multiput (200,248)(3,-3.2){11}{\circle*{0.1}}
\put (230,126) {\line(1,0){20}}
\put (230,216) {\line(1,0){20}}
\put (280,246) {\line(1,0){20}}
\multiput (300,246)(2,-3.2){16}{\circle*{0.1}}
\put (330,198) {\line(1,0){20}}
\thinlines
\put ( 40,265) {$\vdots$} 
\put ( 30,290) {$M_F=2$}
\put ( 80,290) {$M_F=1$}
\put (130,290) {$M_F=3$}
\put (180,290) {$M_F=0$}
\put (230,290) {$M_F=4$}
\put (280,290) {$M_F=1$}
\put (330,290) {$M_F=5$}
\put ( 35, 35) {$K$=0}
\put (115, 35) {$K$=1}
\put (215, 35) {$K$=2}
\put (315, 35) {$K$=3}
\put ( 52, 60) {$0^+_{E}$}
\put ( 52, 90) {$1^+_{E}$}
\put ( 52,150) {$2^+_{E}$}
\put ( 52,240) {$3^+_{E}$}
\put (102, 96) {$1^-_{E}$}
\put (102,156) {$2^-_{E}$}
\put (102,246) {$3^-_{E}$}
\put (152, 80) {$1^-_{A_1 A_2}$}
\put (152,140) {$2^-_{A_2 A_1}$}
\put (152,230) {$3^-_{A_1 A_2}$}
\put (202,158) {$2^+_{A_1 A_2}$}
\put (202,248) {$3^+_{A_2 A_1}$}
\put (252,126) {$2^+_{E}$}
\put (252,216) {$3^+_{E}$}
\put (302,246) {$3^-_{E}$}
\put (352,198) {$3^-_{E}$}
\end{picture}
\caption[Rotational spectrum of $(v_1,v_2^{l=1})$ band]
{\small As Figure~\ref{vib0}, but for a $(v_1,v_2^{l=1})$ vibrational band. 
\normalsize}
\label{vib1}
\vspace{15pt}
\end{figure}

The connection between an algebraic and a geometric description of an 
oblate top can be studied by means of intrinsic (or coherent) states 
\cite{cs}. In such an approach, each vibrational band $(v_1,v_2^l)$ is 
represented by an intrinsic state which can be obtained from 
Eq.~(\ref{cond}) by replacing a condensate boson ($b^{\dagger}_c$) 
by one of the deformed bosons of Eq.~(\ref{bint}). However, 
the deformed operators, $b^{\dagger}_v$ and $b^{\dagger}_w$, do not have 
good projection of the vibrational angular momentum on the symmetry axis. 
In order to construct intrinsic states with well-defined projection on 
the symmetry axis, we transform the cartesian bosons of Eq.~(\ref{bint}) 
to spherical bosons, and introduce the linear combinations \cite{KM}
\ba
\eta^{\dagger}_m &=& ( b^{\dagger}_{\lambda,m} 
+ i \, b^{\dagger}_{\rho,m} )/\sqrt{2} ~,
\nonumber\\
\zeta^{\dagger}_m &=& ( b^{\dagger}_{\lambda,m} 
- i \, b^{\dagger}_{\rho,m} )/\sqrt{2} ~,
\ea
for which we have 
\ba
\eta^{\dagger}_1 &=& 
( - b^{\dagger}_{v} - i \, b^{\dagger}_{w} )/\sqrt{2} ~,
\nonumber\\
\zeta^{\dagger}_{-1} &=& 
( b^{\dagger}_{v} - i \, b^{\dagger}_{w} )/\sqrt{2} ~.
\ea
In this representation, the operator $\hat F_2$ has 
the simple form of the difference between two number operators
\ba
\hat F_2 &=& \hat n_{\zeta} - \hat n_{\eta} ~. 
\ea
The intrinsic state for a vibration $(v_1,v_2^l)$ with 
projection $\pm l$ on the symmetry axis is then 
given by (for $N \rightarrow \infty$)
\ba
| N,v_1,v_2,\pm l;R \rangle &=& 
\frac{( \eta^{\dagger}_{1} )^{(v_2 \pm l)/2}}
{\sqrt{((v_2 \pm l)/2)!}}
\frac{( \zeta^{\dagger}_{-1} )^{(v_2 \mp l)/2}}
{\sqrt{((v_2 \mp l)/2)!}}
\frac{( b^{\dagger}_{u} )^{v_1}}{\sqrt{v_1!}}
\frac{( b^{\dagger}_{c} )^{N-v_1-v_2}}{\sqrt{(N-v_1-v_2)!}} 
\; | 0 \rangle ~, \label{lint}
\ea
which can be expressed in terms of a sum over $(n_{\eta},n_{\zeta})$ 
configurations 
\ba
( \eta_{1}^{\dagger}   )^{(v_2 \pm l)/2}
( \zeta_{-1}^{\dagger} )^{(v_2 \mp l)/2}
( \eta_{-1}^{\dagger}  )^{n_{\eta} -(v_2 \pm l)/2}
( \zeta_{1}^{\dagger}  )^{n_{\zeta}-(v_2 \mp l)/2}
( s^{\dagger}          )^{N-n_{\eta}-n_{\zeta}} \; | 0 \rangle ~.
\ea
For each of these configurations the projection $K$ of 
the angular momentum along the symmetry axis is
\ba
K &=& n_{\zeta}-n_{\eta} \pm 2l \;=\; m_F \pm 2l ~. \label{fkl}
\ea
This shows that the algebraic label $m_F$ has a direct 
interpretation in terms of the geometric labels, $K$ and $l$.
For $l=0$ we have $m_F=K$, but for $l>0$ there are two possible values 
of $m_F$ for each $K$. Thus $M_F=|m_F|=|K \mp 2l|$ provides
an additional quantum number which is needed for a complete 
classification of the rotational excitations of an oblate top. 
For example, the $(v_1,v_2^{l=1})$ vibrational band has two  
$L^{-}_{E}$ levels which have the same value of $K=3$, 
but different values of $M_F$~. The rotational spectrum is given by 
\ba
E_{\mbox{rot}} &=& \kappa_1 \, L(L+1) - \kappa_2 \, M_F^2 
\nonumber\\
&=& \kappa_1 \, L(L+1) - \kappa_2 \, (K^2 \mp 4Kl + 4l^2) ~.
\label{erot}
\ea
The last term contains the effects of the Coriolis force which 
gives rise to a $8 \kappa_2 Kl$ splitting of the rotational 
levels which increases linearly with $K$. 
The label $M_F$ plays a role similar to that of the 
label $G$ discussed by Watson \cite{Watson} for X$_3$ 
molecules, and that of the label $m$ by Bowler {\em et al.} \cite{Hey} 
in the context of a harmonic oscillator quark model in 
baryon spectroscopy.  
As an example of the assignments of $K$ and $M_F$ we
show in Figures~\ref{vib0} and~\ref{vib1} the classification scheme 
for the levels with $L \leq 3$ belonging to a $(v_1,v_2^{l})$ 
vibrational band with $l=0$ and $l=1$, respectively. 

With the exception of the levels with $K=l=0$, all levels in 
Figures~\ref{vib0} and~\ref{vib1} are doubly degenerate. 
According to Eq.(\ref{erot}) the splitting of the levels with $K=0$ 
is zero, whereas for $K>0$ there remains a double degeneracy because 
of the two projections $\pm K$ on the symmetry axis. In particular, 
the rotational spectrum of Figure~\ref{vib1} does not exhibit $l$-type 
doubling, and therefore for $M_F=0$ (mod 3) there is a degenerate 
doublet consisting of $A_1$ and $A_2$ levels. This is a consequence 
of the fact that the one- and two-body $D_3$ invariant Hamiltonians 
of Eqs.~(\ref{ms3}) and~(\ref{oblate}) commute with $\hat F_2$. 
The degeneracy of the $A_1$ and $A_2$ states can be lifted by 
introducing higher order interactions that break $M_F$ symmetry. 
For example, there exist three-body $D_3$ invariant interactions 
that mix states with $\Delta M_F=\pm 6$ 
\cite{BDL}. A similar situation is encountered in Watson's effective 
Hamiltonian \cite{Watson}, whose main terms are diagonal in the 
quantum number $G$ (which plays a similar role as $M_F$), but which 
also contains small higher-order corrections with $\Delta G= \pm 6$~. 

We note that $U(7)$ can also accommodate other types of rotations 
({\em e.g.} nonrigid) and kinematics ({\em e.g.} relativistic). 
In such cases, the $L(L+1)$ and $M_F^2$ terms in Eq.~(\ref{erot}) will 
be replaced by a general function of the angular momentum $L$ and $M_F$ 
\cite{BIL}.

\section{Wave functions}
\setcounter{equation}{0}

In the algebraic approach, the oblate top wave functions 
can be obtained by projection from an intrinsic state.
In principle this is an exact procedure, but since the expressions for 
the intrinsic states of Eq.~(\ref{lint}) are only valid for 
$N \rightarrow \infty$, the same holds for the results presented 
in this section. In the large $N$ limit we find 
\ba
|N,v_1,v_2,l;R;K,L,M \rangle 
&=& \sqrt{\frac{2L+1}{8\pi^2}} \int d\Omega \,
{\cal D}_{MK}^{(L) \, \ast}(\Omega) \, |N,v_1,v_2,l;R,\Omega \rangle ~.
\ea
The angle $\Omega$ specifies the orientation of the intrinsic state 
\ba
|N,v_1,v_2,l;R,\Omega \rangle &=& 
{\cal R}(\Omega) \, |N,v_1,v_2,l;R \rangle ~,
\nonumber\\
{\cal R}(\Omega) &=& \mbox{e}^{-i \psi \hat L_z} \,
\mbox{e}^{-i \theta \hat L_y} \, \mbox{e}^{-i \phi \hat L_z} ~.
\ea
Next we construct states with good $D_3$ symmetry by considering the 
the action of the transposition and the cyclic permutation on the 
projected wave function 
\ba
P(12) \, |N,v_1,v_2,l;R;K,L,M \rangle &=&
(-1)^{v_2+L} \, |N,v_1,v_2,-l;R;-K,L,M \rangle ~,
\nonumber\\
P(123) \, |N,v_1,v_2,l;R;K,L,M \rangle &=&
\mbox{e}^{2 \pi i(K-2l)/3} \, |N,v_1,v_2,l;R;K,L,M \rangle ~.
\label{p12klm}
\ea
Here we take without loss of generality $K \geq 0$ and $l \geq 0$.
States with good $D_3$ symmetry are given by the linear combinations 
\ba
|\psi_1 \rangle &=& \frac{-i}{\sqrt{2(1+\delta_{K,0}\delta_{l,0})}} \,
\left[ |N,v_1,v_2,-l;R;K,L,M \rangle 
-(-1)^{v_2+L} \, |N,v_1,v_2,l;R;-K,L,M \rangle \right] ~,
\nonumber\\
|\psi_2 \rangle &=& 
\frac{(-1)^{\nu}}{\sqrt{2(1+\delta_{K,0}\delta_{l,0})}} \,
\left[ |N,v_1,v_2,-l;R;K,L,M \rangle 
+(-1)^{v_2+L} \, |N,v_1,v_2,l;R;-K,L,M \rangle \right] ~. 
\nonumber\\
\label{wf12}
\ea
These wave functions are characterized by $m_F=K+2l=\nu$ (mod 3).
For $K \neq 0$ and $l \neq 0$ there are two 
extra linear combinations with $m_F=K-2l=\nu$ (mod 3) 
\ba
|\psi_3 \rangle &=& 
\frac{-i}{\sqrt{2(1+\delta_{K,0}\delta_{l,0})}} \,
\left[ |N,v_1,v_2,l;R;K,L,M \rangle 
-(-1)^{v_2+L} \, |N,v_1,v_2,-l;R;-K,L,M \rangle \right] ~,
\nonumber\\
|\psi_4 \rangle &=& 
\frac{(-1)^{\nu}}{\sqrt{2(1+\delta_{K,0}\delta_{l,0})}} \,
\left[ |N,v_1,v_2,l;R;K,L,M \rangle 
+(-1)^{v_2+L} \, |N,v_1,v_2,-l;R;-K,L,M \rangle \right] ~.
\nonumber\\
\label{wf34}
\ea
According to Eq.~(\ref{P123}) the wave functions $| \psi_1 \rangle$ 
($| \psi_2 \rangle$) and $| \psi_3 \rangle$ ($| \psi_4 \rangle$) 
transform for $\nu=0$ as $t=A_2$ ($A_1$), and for 
$\nu=1,2$ as $t=E_{\rho}$ ($E_{\lambda}$). These wave functions are 
consistent with the choice of geometry in the deformed bosons of 
Eqs.~(\ref{bc}) and~(\ref{bint}). 

\section{Baryon resonances}
\setcounter{equation}{0}

\begin{figure}
\centerline{\hbox{
\psfig{figure=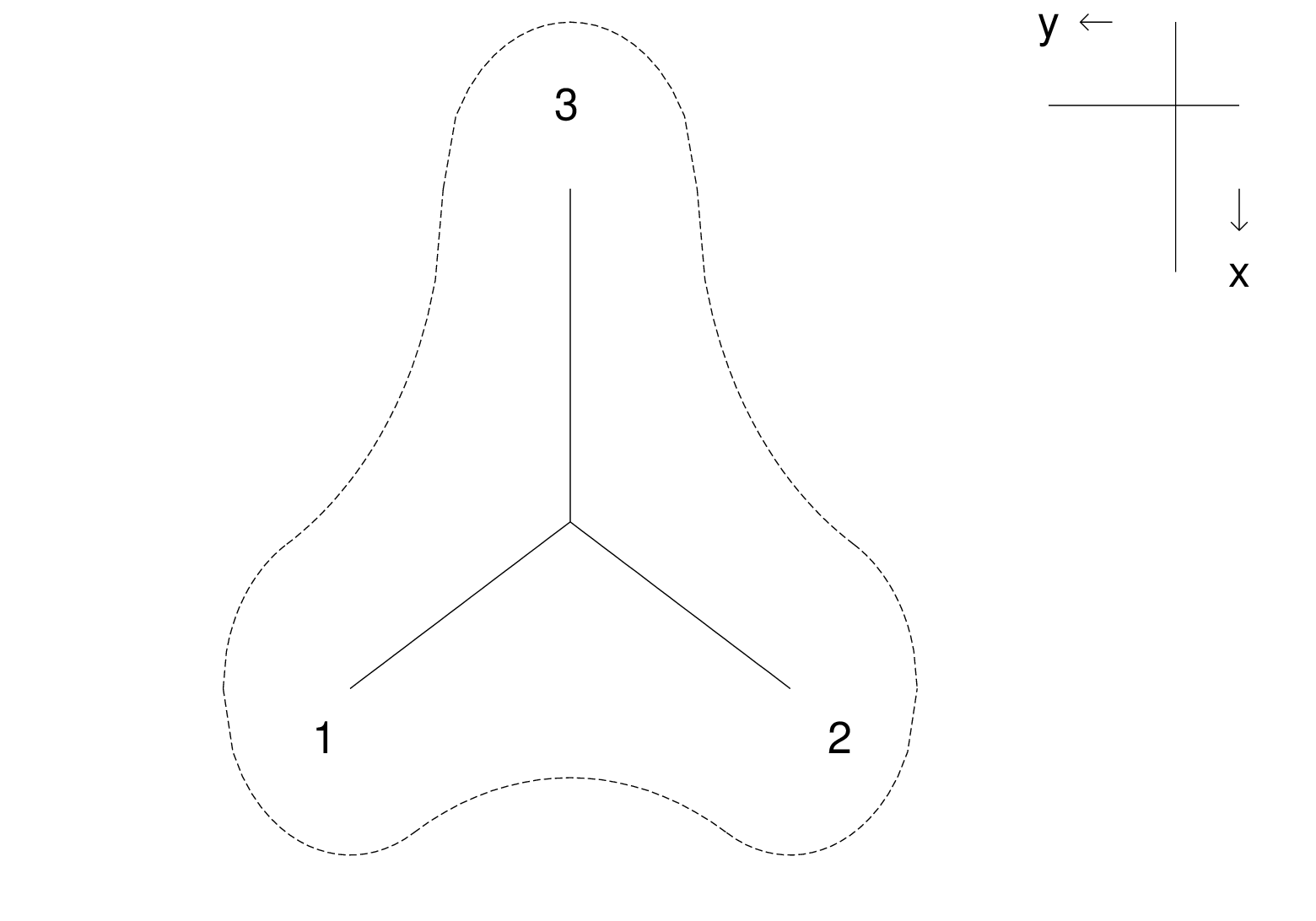,height=0.55\textwidth,width=1.0\textwidth} }}
\caption[Baryon]
{\small Geometry of collective model of baryons.
\normalsize}
\label{geometry}
\vspace{15pt}
\end{figure}

\begin{figure}
\setlength{\unitlength}{1pt}
\begin{picture}(240,200)(-100,0)
\thinlines
\put (  0,  0) {\line(1,0){240}}
\put (  0,  0) {\line(0,1){200}}
\put (  0,200) {\line(1,0){240}}
\put (240,  0) {\line(0,1){200}}
\put (  0, 50) {\line(1,0){5}}
\put (  0,100) {\line(1,0){5}}
\put (  0,150) {\line(1,0){5}}
\put (-75,175) {$M^2$ (GeV$^2$)}
\put (-20, 45) {1}
\put (-20, 95) {2}
\put (-20,145) {3}
\multiput (80,44)(5,0){29}{\circle*{0.1}}
\thicklines
\put ( 20, 44) {\line(1,0){60}}
\put ( 90,104) {\line(1,0){60}}
\put (160,146) {\line(1,0){60}}
\thinlines
\put ( 30, 54) {$N$(939)}
\put (100,114) {$N$(1440)}
\put (170,156) {$N$(1710)}
\put ( 30, 20) {$(0,0^0);0,0^+_{A_1}$}
\put (100, 20) {$(1,0^0);0,0^+_{A_1}$}
\put (170, 20) {$(0,1^1);0,0^+_{E}$}
\put ( 80, 44) {\circle{16}}
\put ( 72, 46) {\vector(0,-1){2}}
\put (120,104) {\vector(0,-1){ 60}}
\put (190,146) {\vector(0,-1){102}}
\end{picture}
\caption[]{\small Schematic representation of the vibrational spectrum
of nucleon resonances. The resonances are labeled by the usual
spectroscopic notation \cite{PDG} and their oblate top classification 
$(v_1,v_2^l);K,L^P_t$~.
\normalsize}
\label{barvib}
\vspace{15pt}
\end{figure}
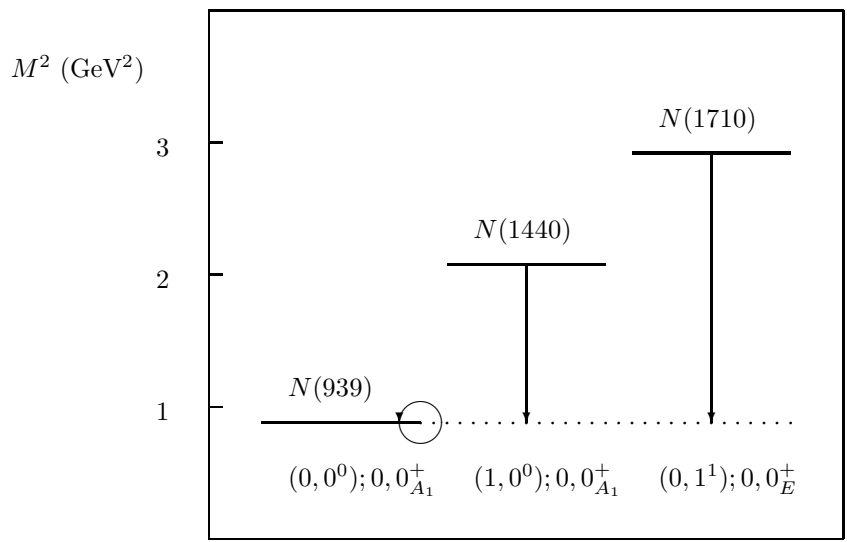

In this section we discuss an application of the oblate top model in 
baryon spectroscopy. We consider baryons to be built of three constituent 
parts (quarks or otherwise) with the string configuration of 
Fig.~\ref{geometry}. 
The full algebraic structure is obtained by combining 
the geometric part, $U(7)$, with the internal spin-flavor-color part, 
$SU_{sf}(6) \otimes SU_c(3)$ (not considering heavy quarks),
\ba
{\cal G} &=& U(7) \otimes SU_{sf}(6) \otimes SU_c(3) ~.
\ea
For the nucleon (isospin $I=1/2$) and delta ($I=3/2$) families of 
resonances the three strings of Fig.~\ref{geometry} have equal lengths 
and equal relative angles. The three constituent parts form 
an equilateral triangle with $D_{3h} \supset D_3$ point group 
symmetry. Baryon resonances are then interpreted in terms of 
vibrations and rotations of an oblate symmetric top. 
In order to have total baryon wave functions that are antisymmetric, 
the permutation symmetry of the geometric (or spatial) part 
must be the same as the permutation symmetry of the spin-flavor part
(the color part is a color singlet, {\it i.e.} antisymmetric). 
Therefore one can also use
the dimension of the $SU_{sf}(6)$ representations to label the states: 
$A_1 \leftrightarrow [56]$, $A_2 \leftrightarrow [20]$ and 
$E   \leftrightarrow [70]$~.
The nucleon itself is identified with the oblate top ground state 
$(v_1,v_2^l);K,L^P_t=(0,0^0);0,0^+_{A_1}$, whereas the $N(1440)$ 
Roper and the $N(1710)$ resonances are interpreted as 
one-phonon excitations with $(1,0^0);0,0^+_{A_1}$ and 
$(0,1^1);0,0^+_{E}$, respectively (see Figure~\ref{barvib}). 

In such a collective model of baryons, the mass operator is written in 
terms of a spatial and a spin-flavor part. The spatial part is given by 
Eq.~(\ref{oblate}) plus a term linear in the angular momentum $L$ 
to reproduce the linear Regge trajectories, while the spin-flavor part 
is given by the G\"ursey-Radicati form \cite{GR}. With this mass operator 
we obtained a good overall fit of the spectrum of the nucleon and delta 
resonances with a r.m.s. deviation of 39 MeV \cite{BIL,BL7}. 

A far more sensitive test of models of baryon structure is provided by 
electromagnetic, strong and weak couplings. 
In the next section we show how the general formalism of the oblate top, 
which was developed in the previous sections, can be used to derive closed 
expressions for helicity amplitudes that can be measured in photo- and 
electroproduction and strong decays of baryon resonances.

\section{Form factors}
\setcounter{equation}{0}

Helicity amplitudes for electromagnetic and strong couplings can
be expressed in terms of some elementary spatial matrix elements 
or form factors. For nonstrange resonances these are the matrix 
elements of the operators \cite{BIL,emff,strong}
\ba
\hat U &=& \mbox{e}^{ -i k \beta \hat D_{\lambda,z}/X_D } ~,
\nonumber\\
\hat T_{\lambda,m} &=& 
\frac{i m_3 k_0 \beta}{2 X_D} \left( \hat D_{\lambda,m} \, 
\mbox{e}^{ -i k \beta \hat D_{\lambda,z}/X_D } +
\mbox{e}^{ -i k \beta \hat D_{\lambda,z}/X_D } \, \hat D_{\lambda,m}  
\right) ~, 
\nonumber\\
\hat T_{\rho,m} &=& 
\frac{i m_3 k_0 \beta}{2 X_D} \left( \hat D_{\rho,m} \, 
\mbox{e}^{ -i k \beta \hat D_{\lambda,z}/X_D } +
\mbox{e}^{ -i k \beta \hat D_{\lambda,z}/X_D } \, \hat D_{\rho,m}  
\right) ~, \label{emop}
\ea
with ($m=-1,0,1$). Here $\vec{k}=k \hat z$ is the photon (meson) momentum, 
$k_0$ is the photon (meson) energy, $m_3$ is the 
constituent mass and $\beta$ represents the scale of the coordinate.
The operators $\hat D_{\rho,m}$ and $\hat D_{\lambda,m}$ 
are dipole operators in $U(7)$
\ba
\hat D_{\rho,m} &=& (b^{\dagger}_{\rho} \times \tilde{s} -
s^{\dagger} \times \tilde{b}_{\rho})^{(1)}_m ~, 
\nonumber\\
\hat D_{\lambda,m} &=& (b^{\dagger}_{\lambda} \times \tilde{s} -
s^{\dagger} \times \tilde{b}_{\lambda})^{(1)}_m ~. \label{dipole}
\ea
The normalization factor $X_D$ is given by the reduced matrix element 
between the $L^P_t=0^+_{A_1}$ ground state and the first excited 
$1^-_{E}$ state (both belonging to the $(v_1,v_2^l)=(0,0^0)$ ground 
band) 
\ba
X_D &=& \langle 1^-_E || \hat D_{\lambda} || 0^+_{A_1} \rangle
  \;=\; -\langle 1^-_E || \hat D_{\rho}    || 0^+_{A_1} \rangle ~. 
\ea
In the large $N$ limit (and $R^2 > 0$) it reduces to 
\ba
\lim_{N \rightarrow \infty} X_D &=& \frac{NR\sqrt{2}}{1+R^2} ~.
\label{xd}
\ea
Since $\hat D_{\lambda}$ is a generator of the algebra of $U(7)$, 
the matrix elements of $\hat U$ are representation 
matrix elements of $U(7)$, {\it i.e.} a generalization of the Wigner 
${\cal D}$-matrices for $SU(2)$. By making an appropriate basis 
transformation they can be obtained numerically without having to 
make any further approximations. However, in the 
limit of $N \rightarrow \infty$ (infinitely large model space)
the matrix elements of $\hat U$, $\hat T_{\lambda,m}$ and 
$\hat T_{\rho,m}$ can also be derived in closed form. 

We illustrate the method by evaluating the matrix elements of 
$\hat U$ that connect the nucleon with its excited rotational states. 
In the collective model, the nucleon wave function is that of the ground
state of the oblate symmetric top 
\ba
|\psi_0 \rangle &=& |N,v_1=0,v_2=0,l=0;R;K=0,L=0,M=0 \rangle ~. 
\label{nucleon}
\ea
The vibrationally elastic matrix element connecting the ground state 
with a rotational excitation is given by
\ba
&& \langle N,0,0,0;R;K,L,M | \, \hat U \, | N,0,0,0;R;0,0,0 \rangle 
\nonumber\\
&& \hspace{1cm} \;=\; \frac{\sqrt{2L+1}}{8\pi^2} 
\int d \Omega \, d \Omega^{\prime} \; {\cal D}^{(L)}_{MK}(\Omega) \, 
{\cal D}^{(0) \, \ast}_{00}(\Omega^{\prime}) \, 
\langle N,0,0,0;R,\Omega | \, \hat U \, | 
N,0,0,0;R,\Omega^{\prime} \rangle ~.
\ea
In the large $N$ limit the matrix element appearing in the
integrand becomes diagonal in the orientation $\Omega$ of the condensate. 
The remaining angular integral can be obtained in closed form in 
terms of a spherical Bessel function 
\ba
&& \frac{\sqrt{2L+1}}{8\pi^2} 
\int d \Omega \; {\cal D}^{(L)}_{MK}(\Omega) \, 
\mbox{e}^{i \, k\beta \sin \theta \cos \phi} 
\nonumber\\
&& \hspace{1cm} \;=\; \delta_{M,0} \, i^K \, \sqrt{2L+1} \,
\frac{\sqrt{(L+K)!(L-K)!}}{(L+K)!!(L-K)!!} \,
\frac{1}{2} [ 1+(-1)^{L-K} ] \, j_L(k\beta) ~. \label{klm}
\ea
With the wave functions of Eq.~(\ref{wf12}) we find that 
the matrix elements connecting the nucleon with its 
rotationally excited states are  
\ba
\langle \psi_1 | \, \hat U \, | \psi_0 \rangle &=& 0 ~,
\nonumber\\
\langle \psi_2 | \, \hat U \, | \psi_0 \rangle 
&=& \delta_{M,0} \, 
(-1)^{\nu} \, i^K \, \sqrt{\frac{2L+1}{2(1+\delta_{K,0})}} \,
\frac{\sqrt{(L+K)!(L-K)!}}{(L+K)!!(L-K)!!} \, [1+(-1)^{L-K}] \, 
j_L(k\beta) ~. \label{rotme}
\ea
For the $(v_1,v_2^l)=(0,0^0)$ ground band we have $K=\nu$ (mod 3). 

To summarize the results for the matrix elements of Eq.~(\ref{emop}) 
we introduce the notation
\ba
F_{i}(k) &=& 
\langle \psi_i | \, \hat U \, | \psi_0 \rangle ~,
\nonumber\\
G_{i;\lambda,m}(k) &=& 
\langle \psi_i | \, \hat T_{\lambda,m} \, | \psi_0 \rangle ~,
\nonumber\\
G_{i;\rho,m}(k) &=& 
\langle \psi_i | \, \hat T_{\rho,m} \, | \psi_0 \rangle ~, 
\ea
for the elementary oblate top form factors. The nucleon wave function 
$|\psi_0 \rangle$ is given in Eq.~(\ref{nucleon}), and the wave function 
of the resonance $|\psi_i \rangle$ is given in Eq.~(\ref{wf12}) for 
$i=1,2$ and in Eq.~(\ref{wf34}) for $i=3,4$~. All matrix elements 
can be expressed in terms of spherical Bessel functions. 
For the rotational transitions we find 
\ba
F_1(k) &=& 0 ~,
\nonumber\\
F_2(k) &=& \delta_{M,0} \, Z(k \beta) ~, 
\ea
where $Z(k \beta)$ can be obtained by comparing with Eq.~(\ref{rotme}).
The matrix elements of $\hat T_{\lambda,m}$ and 
$\hat T_{\rho,m}$ can be derived as
\ba
G_{1;\lambda,m}(k) &=& 0 ~,
\nonumber\\
G_{2;\lambda,z}(k) &=& - \delta_{M,0} \, m_3 k_0 \beta \,
\frac{d Z(k \beta)}{d k \beta} ~,
\nonumber\\
G_{2;\lambda,\pm}(k) &=& \pm \delta_{M,\pm 1} \, m_3 k_0 \beta \, 
\sqrt{L(L+1)} \, \frac{Z(k \beta)}{k \beta} ~, 
\label{glambda}
\ea
and
\ba
G_{1;\rho,z}(k) &=& -\delta_{M,0} \, m_3 k_0 \beta \, 
K(-1)^{\nu} \, \frac{Z(k \beta)}{k \beta} ~,
\nonumber\\
G_{1;\rho,\pm}(k) &=& \pm \delta_{M,\pm 1} \, m_3 k_0 \beta \, i^{K} \, 
\sqrt{\frac{2L+1}{2(1+\delta_{K,0})}} \, 
\frac{\sqrt{(L+K)!(L-K)!}}{(L+K)!!(L-K)!!} \,
\frac{K}{\sqrt{L(L+1)}}
\nonumber\\
&& \times \left[ 1+(-1)^{L-K} \right] \, 
\frac{1}{2L+1} \, \left[ (L+1) \, j_{L-1}(k\beta) 
- L \, j_{L+1}(k\beta) \right] 
\nonumber\\
&& +\delta_{M,\pm 1} \, m_3 k_0 \beta \, i^{K} \, 
\sqrt{\frac{2L+1}{2(1+\delta_{K,0})}} \, 
\frac{\sqrt{(L+K)!(L-K)!}}{(L+K-1)!!(L-K-1)!!} \, 
\frac{1}{\sqrt{L(L+1)}}
\nonumber\\
&& \times \left[ 1-(-1)^{L-K} \right] \, j_{L}(k\beta) ~,
\nonumber\\
G_{2;\rho,m}(k) &=& 0 ~.
\label{grho} 
\ea
In Eqs.~(\ref{glambda}) and~(\ref{grho}) we have used 
$G_{.;.,\pm}(k)=\mp \sqrt{2} \, G_{.;.,\pm 1}(k)$. 
We see that all relevant matrix elements can be expressed in terms
of $F_2(k)$ and $G_{1;\rho,\pm}(k)$ only. In Table~\ref{otff1} we present 
the results for the elementary form factors that are relevant for the 
lowlying nucleon and delta resonances with $L \leq 2$.

The matrix elements of Eq.~(\ref{emop}) for vibrational 
excitations are obtained as above by projection from the 
corresponding intrinsic states. In Table~\ref{otff2} we present the 
results for the elementary form factors for the vibrational excitations 
$(0,0^0) \rightarrow (1,0^0)$, $(0,1^1)$ for nucleon and 
delta resonances with $L \leq 1$. 
The oblate top form factors in Tables~\ref{otff1} and~\ref{otff2} 
correspond to the transitions indicated schematically in 
Figure~\ref{barvib}.

\section{Summary and conclusions}
\setcounter{equation}{0}

In this contribution, we presented an algebraic 
treatment of the three-body problem. We used the method of bosonic 
quantization, which for the two relative coordinates of the three-body 
system gives rise to a $U(7)$ spectrum generating algebra. The model 
space is spanned by the symmetric irreps $[N]$ of $U(7)$.

In particular, we studied the case of three identical objects and 
showed how the corresponding permutation symmetry can be taken into 
account exactly. Rather than explicity constructing states of good 
permutation symmetry \cite{KM}, we obtain these states by diagonalizing 
a Hamiltonian that is invariant with respect to the point group symmetry. 
We developed a general procedure to determine the permutation symmetry 
of any given wave function. It was shown that in the large $N$ limit 
(large model space) the algebraic Hamiltonian for the X$_3$ system 
corresponds to an oblate symmetric top. 

For the special case of one- and two-body interactions, the 
eigenstates can be labeled by an additional quantum number $M_F$. 
This label plays a very interesting role. On the one hand, it has a 
direct connection to the permutation symmetry. On the other hand,  
in the large $N$ limit it is directly related to the geometric labels 
$K$ and $l$, and provides an extra label which is needed to classify 
the rovibrational states of the oblate top uniquely. 

It was shown that $U(7)$ provides a unified treatment of 
both rotational and vibrational excitations of an oblate top.
The ensuing algebraic treatment of the oblate top has found useful 
applications both in molecular physics (X$_3$ molecules
\cite{BDL,Frank}) and hadron physics (nonstrange qqq baryons
\cite{BIL,emff,strong}). As an example,  
we discussed an application to baryon resonances 
which are interpreted as vibrations and rotations of an oblate top. 
We derived closed expressions for a set of elementary form factors 
for nonstrange resonances. These elementary form factors appear in 
all calculations of electromagnetic, strong and weak couplings, 
and hence form the backbone of the model. In a separate contribution 
to these proceedings \cite{LB} we discuss in more detail the 
application to electromagnetic and strong couplings of baryon resonances. 

Finally, we note that, although we discussed the specific case of 
three identical objects, the algebraic procedure is valid in general, 
both for three nonidentical objects and for the many-body system. 
Applications of these techniques to polyatomic molecules 
will be discussed in a separate contribution \cite{AF}.

\section*{Acknowledgements}

It is a pleasure to thank A. Frank for interesting discussions 
and a careful reading of the manuscript. 
This work is supported in part by CONACyT, M\'exico under project 
400340-5-3401E and DGAPA-UNAM under project IN105194 (R.B.), 
and by grant No. 94-00059 from the United States-Israel Binational 
Science Foundation (BSF), Jerusalem, Israel (A.L.).

\clearpage

\clearpage

\clearpage

\clearpage

\clearpage

\begin{table}
\centering
\caption[$\Delta(v_1+v_2)=0$ oblate top form factors]{\small 
Analytic expressions of the $(v_1,v_2^l)=(0,0^0) \rightarrow (0,0^0)$ 
matrix elements of the transition operators of Eq.~(\ref{emop}) in the 
oblate top model for $N \rightarrow \infty$. 
The final states are labeled by 
$[\mbox{dim}\{SU_{sf}(6)\},L^P]_{(v_1,v_2^l);K}$~.
The initial state is $[56,0^+]_{(0,0^0);0}$.
\normalsize}
\label{otff1} \vspace{15pt} 
\begin{tabular}{ccc}
\hline
& & \\
Final state & $F_2(k)$ & $G_{1;\rho,\pm}(k)/m_3 k_0 \beta$ \\
& & \\
\hline
& & \\
$[56,0^+]_{(0,0^0);0}$ & $j_0(k \beta)$ & 0 \\
& & \\
$[20,1^+]_{(0,0^0);0}$ & 0 
& $\frac{\sqrt{3}}{\sqrt{2}} \, j_1(k \beta)$ \\
& & \\
$[70,1^-]_{(0,0^0);1}$ & $-i \, \sqrt{3} \, j_1(k \beta)$ 
& $\pm i \, \frac{1}{\sqrt{6}} \, [2j_0(k \beta)-j_2(k \beta)]$ \\
& & \\
$[56,2^+]_{(0,0^0);0}$ & $\frac{\sqrt{5}}{2} \, j_2(k \beta)$ & 0 \\
& & \\
$[70,2^-]_{(0,0^0);1}$ & 0 
& $i \, \frac{\sqrt{5}}{\sqrt{2}} \, j_2(k \beta)$ \\ 
& & \\
$[70,2^+]_{(0,0^0);2}$ & $-\frac{\sqrt{15}}{2} \, j_2(k \beta)$ 
& $\mp \frac{1}{\sqrt{10}} \, [3j_1(k \beta)-2j_3(k \beta)]$ \\
& & \\
\hline
\end{tabular}
\vspace{15pt}
\end{table}

\clearpage

\begin{table}
\centering
\caption[$\Delta(v_1+v_2)=1$ oblate top form factors]{\small 
As Table~\ref{otff2}, but for the 
$(v_1,v_2^l)=(0,0^0) \rightarrow (1,0^0)$ and $(0,1^1)$ transitions.
$\chi_1=(1-R^2)/2R\sqrt{N}$ and $\chi_2=-\sqrt{1+R^2}/R\sqrt{2N}$~.
\normalsize}
\label{otff2} \vspace{15pt} 
\begin{tabular}{ccc}
\hline
& & \\
Final state & $F_2(k)$ & $G_{1;\rho,\pm}(k)/m_3 k_0 \beta$ \\
& & \\
\hline
& & \\
$[56,0^+]_{(1,0^0);0}$ & $-\chi_1 \, k \beta \, j_1(k \beta)$ & 0 \\
& & \\
$[20,1^+]_{(1,0^0);0}$ & 0 
& $\chi_1 \, \frac{1}{\sqrt{6}} \, 
k \beta \, [2j_0(k \beta)-j_2(k \beta)]$ \\
& & \\
$[70,1^-]_{(1,0^0);1}$ & $-i \, \chi_1 \, 
\frac{1}{\sqrt{3}} \, k \beta \, [j_0(k \beta) - 2j_2(k \beta)]$ 
& $\pm i \, \chi_1 \, \frac{1}{\sqrt{6}} \, 
[2j_0(k \beta)+2j_2(k \beta) - 3 k \beta \, j_1(k \beta)]$ \\
& & \\
\hline
& & \\
$[70,0^+]_{(0,1^1);0}$ & $-\chi_2 \, \frac{1}{\sqrt{2}} \,
k \beta \, j_1(k \beta)$ & 0 \\
& & \\
$[70,1^+]_{(0,1^1);0}$ & 0 
& $-\chi_2 \, \frac{\sqrt{3}}{2} \, k \beta \, j_2(k \beta)$ \\
& & \\
$[56,1^-]_{(0,1^1);1}$ & $-i \, \chi_2 \, 
\frac{\sqrt{3}}{2} \, k \beta \, j_2(k \beta)$ & 0 \\
& & \\
$[20,1^-]_{(0,1^1);1}$ & 0 & $\pm i \, \chi_2 \, 
\frac{\sqrt{3}}{2\sqrt{2}} \, [3j_2(k \beta) - k \beta \, j_1(k \beta)]$ \\
& & \\ 
\hline
& & \\
& $F_4(k)$ & $G_{3;\rho,\pm}(k)/m_3 k_0 \beta$ \\
& & \\
\hline
& & \\
$[70,1^-]_{(0,1^1);1}$ & $-i \, \chi_2 \, 
\frac{1}{2\sqrt{3}} \, k \beta \, [2j_0(k \beta) - j_2(k \beta)]$ 
& $\pm i \, \chi_2 \, \frac{1}{2\sqrt{6}} \, 
[j_2(k \beta)+4j_0(k \beta)+3 k\beta \, j_1(k \beta)]$ \\
& & \\
\hline
\end{tabular}
\vspace{5pt}
\end{table}

\end{document}